\documentclass[twocolumn,showpacs,amsmath,amssymb,pra,superscriptaddress,floatfix]{revtex4}

\usepackage{graphicx}% Include figure files
\usepackage{dcolumn}% Align table columns on decimal point
\usepackage{bm}% bold math
\usepackage{subfigure}

\newcommand{\tn}[1]{\textnormal{#1}}

\begin{document}

%\preprint{APS/123-QED}

\title{Resonances of the helium atom in a strong magnetic field}
\date{\today}

\author{Armin L\"uhr}
\email[Present\ address:\ Institut f\"ur Physik, Humboldt-Universit\"at zu Berlin, Hausvogteiplatz 5-7, D-10117 Berlin, Germany; Armin.Luehr@physik.hu-berlin.de]{}
\affiliation{%
Theoretische Chemie, Institut f\"ur Physikalische Chemie,
Universit\"at Heidelberg,
INF 229, 69120 Heidelberg, Germany}%
\author{Omar-Alexander Al-Hujaj}
%\email[ ]{ }
\affiliation{%
Theoretische Chemie, Institut f\"ur Physikalische Chemie,
Universit\"at Heidelberg,
INF 229, 69120 Heidelberg, Germany}%
\author{Peter Schmelcher}
\email[]{Peter.Schmelcher@pci.uni-heidelberg.de}
\affiliation{%
Theoretische Chemie, Institut f\"ur Physikalische Chemie,
Universit\"at Heidelberg,
INF 229, 69120 Heidelberg, Germany}%
\affiliation{%
Physikalisches Institut, Universit\"at Heidelberg, 
Philosophenweg 12, 69120 Heidelberg, Germany}%

\date{\today}% It is always \today, today,

\begin{abstract}\label{txt:abstract}
We present an investigation on the resonances of a doubly excited helium atom 
in a strong magnetic field covering the regime $B=0-100$ a.u. A 
full-interaction approach which is based on an anisotropic Gaussian basis 
set of one-particle functions being nonlinearly optimized for each field strength is
employed. Accurate results for a total of 17 resonances below the threshold of 
consisting of He$^+$ in the $N=2$ state are reported in this work. This includes states with 
total magnetic quantum numbers $M=0,-1,-2$ and even $z$-parity. The corresponding
binding energies 
are compared to approximate energies of two-particle configurations consisting of 
two hydrogen-like electrons in the strong field regime, thereby providing an 
understanding of the behavior of the energies of the resonances with 
varying field strength.
\end{abstract}

\maketitle
%%%%%%%%%%%%%%%%%%%%%%%%%%%%%%%%%%%%%%
%
\section{Introduction}\label{sec:introduction}
%
%%%%%%%%%%%%%%%%%%%%%%%%%%%%%%%%%%%%%%
During the last years the number of observed magnetic white dwarfs (MWD) 
with magnetic field strengths $B$ between a few and 10$^5$~Tesla has grown rapidly. 
Including the ESO~\cite{reimers} and SDSS~\cite{schmidt,vanlandingham} surveys 
there are about 170 MWD detected~\cite{euchner} and cataloged
\footnote{http://www.astronomy.villanova.edu/WDCatalog/index.html}. 
Even stronger fields have been found in the vicinity of neutron stars 
($B\approx10^7-10^9$~T)~\cite{trumper}.  
Therefore, the behavior and properties of matter exposed to strong magnetic 
fields is of immediate interest. In this context, atomic physics plays a 
major role with respect to analyzing the atmospheres of these astronomical objects. 
Spectra resulting from theoretical investigations are a key ingredient in order to 
identify atomic (or molecular) species in the atmospheres of MWD. 
Beyond this they allow to gain information on the magnetic field strength 
and its landscape of the object under investigation.

Most of the literature on atoms in strong magnetic fields deals with the 
hydrogen atom which has also become a paradigm for a low-dimensional chaotic 
system~\cite{friedrich}. Already in the 1930s it was studied employing 
perturbation theory in the low field regime ($B\lesssim 2\cdot10^3$~T)~\cite{schiff} 
and later on also in the high field regime.
But it was not until 1984 that exact energies of the hydrogen atom for the complete 
magnetic field range $0\leq B\lesssim 5\cdot 10^8$~T were 
available~\cite{forster,henry,ruder}. The calculation of accurate extensive 
data was made possible by the advent of supercomputers. It was the accuracy 
and completeness of this new quantum mechanical data which permitted, after 
40 years of speculation, an understanding of the mysterious absorption 
features in the spectrum of the MWD Grw+70$^\circ$8247. They could be 
explained in 1985 as a hydrogen spectrum in a magnetic field with a 
polar value of $3.2\cdot 10^4$~T~\cite{angle,greenstein,wickramasinghe1}. 
In the following decades a great number of MWD with the same type of spectrum have
been identified ~\cite{wickramasinghe1}.

In spite of this great success, there were also spectra of MWD which could not 
be explained in terms of hydrogen lines. Therefore, also results of atoms 
with more than one electron are needed. Ten years ago the knowledge about 
atoms with more than one electron in strong magnetic fields was still very sparse. 
Even in the case of two-electron atoms the investigations covered only 
much smaller parts of the spectrum compared to the case of the hydrogen atom, 
and their accuracy was considerably poorer. At the end of the 1990s, 
sufficiently accurate and extensive data for the helium atom in the magnetic 
field regime $0\leq B \leq 2.35 \cdot 10^7$~T~\cite{becken1,becken2,becken3,jordan1} 
and later also for $2.35 \cdot 10^7$~T $\leq B \leq 2.35 \cdot 10^9$~T~\cite{alex1} 
emerged. These data consist exclusively of bound-state properties 
and the corresponding transitions.

Employing these data, it is possible to identify astronomical spectra 
of strongly magnetized helium atmospheres. In particular the interpretation 
of the absorption features in the spectrum of the MWD GD~229, one of the major 
remaining mysteries for more than 25 years, could be solved. It was the first high-field 
MWD with an identified helium atmosphere ~\cite{jordan2}.

Nevertheless, the spectra of several MWD with an estimated hydrogen or 
helium atmosphere show major absorption features that remain unexplained. 
There were indications to explain the unresolved features with bound-free 
and free-free transitions but also with spectral components of heavier elements 
or of ions. However, the literature concerning these topics is comparably sparse. 
In order to shed light on this issue there were efforts to calculate 
bound-free opacities for the hydrogen atom~\cite{merani}. More recently, 
extensive and accurate calculations on the bound state spectrum of heavier 
elements like lithium~\cite{alex2} and beryllium~\cite{alex3} in strong 
magnetic fields have been accomplished. However, these newer results 
have not been able to explain the unresolved absorption features in question. 
 
In view of the above it is of particular interest to explore the doubly 
excited states, i.e., the resonances of helium in strong magnetic fields, 
which have not been considered to date in the literature. It is the 
aim of this work to provide a step towards the understanding of doubly 
excited helium atoms in strong magnetic fields and to provide the 
first data for the corresponding resonances.

The paper is organized as follows: In Section II we introduce the 
Hamiltonian and discuss its symmetries. In Sec. III we report on 
our computational approach and the employed optimized basis set. 
Sec. IV begins with a comparison of calculated resonance energies 
with the  literature in the well-known field-free case. 
Subsequently the resonance energies and their behavior for different magnetic 
field strengths are presented and discussed. The dominant configurations 
contributing to the resonances are identified allowing us to interprete them. 
Additionally, the widths of selected 
resonances are studied with increasing magnetic field strength.

%%%%%%%%%%%%%%%%%%%%%%%%%%%%%%%%%%%%%%
%
\section{Hamiltonian and Symmetries}\label{sec:hamiltonian}
%
%%%%%%%%%%%%%%%%%%%%%%%%%%%%%%%%%%%%%%

Our investigation is carried out under the assumption of an infinite nuclear mass. 
However, by exploiting a pseudo-separation of the center-of-mass motion in a 
magnetic field~\cite{johnson,schmelcher3,schmelcher1} a scaling law can be derived 
which allows one to translate the data for fixed nucleus into results with a 
finite nuclear mass under the assumption of a vanishing pseudomomentum 
(see ref.~\cite{becken1} and refs. therein). The finite nuclear mass effects 
have certainly to be taken into account in the high field regime~\cite{alex1}. 
However, in this work we focus on the intermediate field regime and can therefore 
safely neglect the corresponding considerations.

Assuming the magnetic field $\bm{B}$ to point in the positive $z$-direction 
and for the field strength $B=\left|\bm{B}\right|$, our Hamiltonian reads 
(atomic units are used throughout the paper)
\begin{eqnarray}\label{hamiltonian1}
H &= & \sum_{i=1}^2  (\,\frac{1}{2}\bm{p}_{i}^2-\frac{2}{|\bm{r}_{i}|}+
\frac{1}{2}B(l_i)_z+ \frac{B^2}{8}(x_{i}^2+y_{i}^2)  \\ 
& &\qquad \left. +\, B(s_i)_{z}\,\right)\ +\, \frac{1}{|\bm{r}_2 -\bm{r}_1 |}. \nonumber
\end{eqnarray}
The one-particle operators in equation (\ref{hamiltonian1}) are the 
field-free kinetic energies $\frac{\bm{p}^2_i}{2}$, the Coulomb potential 
energies $-\frac{2}{|\bm{r}_{i}|}$ of the electrons in the field of the nucleus, 
the Zeeman terms $\frac{1}{2}B(l_i)_z$, the diamagnetic terms 
$\frac{B^2}{8}(x_{i}^2+y_{i}^2)$ and the spin energies $B(s_i)_{z}$. The 
two-particle operator $\frac{1}{|\bm{r}_2 -\bm{r}_1 |}$ represents the 
electron-electron repulsion energy. For remarks on the influence of 
relativistic effects and on the electron spin $g$-factor, here taken to 
be equal to 2, we refer the reader to~\cite{ruder, becken1}.

There exist four independent quantities commuting with the 
Hamiltonian (\ref{hamiltonian1})%~\cite{becken1}: The total spin $\bm S^2$, 
the $z$-component $S_z$ of the total spin, the $z$-component $L_z$ of the 
total orbital angular momentum and the total spatial $z$-parity $\Pi_z$. 
Therefore, the following investigations take place in a subspace of a 
specified symmetry, i.e. with given eigenvalues of $\bm{S}^2$, $S_z$, 
$L_z$ and $\Pi_z$. Furthermore, we will denote the calculated states by 
the \emph{magnetic field spectroscopic notation}
\begin{equation}\label{field-notation}
_N\,\nu\ ^{2S+1}\ M\ ^{\Pi_z} ,
\end{equation}
where $2S+1$ is the spin multiplicity, $M$ the eigenvalue of $L_z$ 
and $\nu=1,2,3\ldots$ denotes the degree of excitation within a given 
symmetry subspace, whereas $N$ represents the excitation of the "inner" 
electron. The latter is not an exact quantum number but a helpful 
spectroscopic label within an independent-particle picture.

%%%%%%%%%%%%%%%%%%%%%%%%%%%%%%%%%%%%%%
%
\section{Computational Approach}\label{sec:computation}
%
%%%%%%%%%%%%%%%%%%%%%%%%%%%%%%%%%%%%%%

The Schr\"odinger equation is solved by applying a full configuration-interaction 
(full CI) approach. For our calculations we use an anisotropic 
Gaussian basis set which has been put forward by Schmelcher and 
Cederbaum~\cite{schmelcher2} for the purpose of investigating atoms and 
molecules in strong magnetic fields. This one-particle basis set has been 
successfully applied to several atoms, ions and 
molecules~\cite{becken1,becken2,becken3,alex1,alex2,alex3,kappes,detmer,alex4,alex5}. 
The basis functions are optimized for each field strength and each 
symmetry separately using a nonlinear optimization procedure described in 
efs.~\cite{becken1,alex1}. A two-particle basis set within a subspace 
($M,\Pi_z$) is constructed by selecting configurations of two optimized 
one-particle orbitals respecting the symmetries. The full CI approach leads 
to a generalized eigenvalue problem in which all matrix elements can be 
calculated analytically and evaluated efficiently~\cite{becken1,becken2}.
Further technical aspects are discussed in refs.~\cite{becken1,alex1,alex4}.

For the investigation in this paper we use the complex-scaling or 
complex-rotation method. This means that all spatial coordinates $\bm{r}$ 
in the Hamiltonian are formally replaced with $\bm{r}e^{i\theta}$ 
(cf.~\cite{reinhardt,moiseyev1}). This leads to a non-Hermitian 
Hamiltonian with complex eigenvalues $W=E-\frac{i}{2}\Gamma$ with the 
energy $E$ and resonance width $\Gamma$ and to a complex-symmetric 
matrix representation. The energies of the bound states 
are unaffected by this transformation, while the different continua are 
rotated about their respective thresholds by the angle $-2\theta$ from 
the real axis into the complex energy plane. The hidden resonances 
of the Hamiltonian in the continuum, associated with complex 
eigenvalues $W$, are revealed if a sufficiently large value of $\theta$ is chosen.

It is one major advantage of the complex-scaling method that the 
sophisticated numerical code developed for bound atom systems 
can be directly 'translated' to the case of unbound resonances. 
However, the complex eigenvalues $W$ of the resonances are only 
independent of the rotation angle $\theta$ if the basis set is 
complete, which is not the case in any approximate numerical investigation. 
Therefore, the rotated Hamiltonian has to be calculated for different values 
of $\theta$. The complex variational theorem and the complex 
Hellmann-Feynman theorem provide a criterion for choosing an 
optimum resonance parameter $\theta$ within a given basis set~\cite{moiseyev2}:
\begin{equation}\label{theta-opt}
\left . \frac{dW(\theta)}{d\theta}\right|_{\theta_\tn{opt}}=0 .
\end{equation} 
To find that optimum $\theta_\tn{opt} \in \mathbb{C}$, one thus has to vary the imaginary part of $\theta$, too. This can be 
interpreted as a real scaling parameter of the wave functions. 
Obviously, the criterion (\ref{theta-opt}) is only fulfilled approximately 
since we can only calculate finite differences between eigenvalues belonging to 
Hamiltonians with a different value of the rotation angle $\theta$.

In our investigation the diagonalization of the rotated Hamiltonian 
is carried out for about 50 to 80 different values of $\theta$. 
The used basis-set size varies from 4000 to 6000 two-particle configurations. 
This results in a computational effort of approximately 30 to 200 hours CPU time 
on a powerful PC \textit{for each magnetic field strength and symmetry subspace}.

%\clearpage

%%%%%%%%%%%%%%%%%%%%%%%%%%%%%%%%%%%%%%
%
\section{Results}\label{sec:results}
%
%%%%%%%%%%%%%%%%%%%%%%%%%%%%%%%%%%%%%%
 
In principle, one distinguishes two kinds of excitations of the helium atom 
using the independent-particle picture. On the one hand, there are the 
\emph{single excitations} of the helium atom. Their "inner" electron occupies 
predominantly the one-particle ground state of He$^+$. The second "outer" 
electron carries, like a hydrogenic electron, the angular momentum alone. 
All bound states are of this kind. On the other hand, there are the 
\emph{doubly excited} helium states where the "inner" electron is excited, too. 
They appear as resonances in the continuum of the singly excited helium states. 
The degree of the "inner" and "outer" excitation is denoted by $N$ with 
$N=1,2,\ldots $ and by $n$ with $n\geq N$, respectively.
States with the same "inner" excitation $N$ form Rydberg series converging 
energetically to the threshold $I_N$. These thresholds correspond to the 
energies of an excited helium ion He$^+$ in the $N$-th energy level.

\subsection{Resonances in the field-free case} 

Doubly excited (DE) helium has been an active research topic for the last 40 years, 
both from the numerical and analytical standpoint (cf. review~\cite{tanner}). Major contributions 
to the calculation of the energy levels were provided by, e.g., B\"urgers et al. (S-states) 
\cite{burgers}, Ho et al. (P- and D-states) \cite{hoP,hoD} or 
Bachau ($N=3$ states) \cite{bachau}. However, the data have not reached the same 
accuracy and completeness compared with the literature of singly excited helium. 
DE helium is considered as a standard model for a strongly correlated system. 
We use the standard notation i.e. the \emph{correlation quantum numbers} introduced by 
Herrick, Sinano$\mathrm{\breve{g}}$lu~\cite{herrick} and Lin~\cite{lin}:
\begin{equation}\label{free-notation}
^{2S+1} L^\Pi \,_N(K,T)^A_n\,.
\end{equation}
Here $L$ is the total angular momentum, and $K$ and $T$ represent 
angular and $A$ radial correlations. 

However, the focus of this paper lies on the interaction of the atomic system 
with a strong magnetic field. For a finite field strength the correlation 
quantum numbers (\ref{free-notation}) lose their meaning. Therefore, it is 
instructive to consider the correspondence between the magnetic field 
spectroscopic notation~(\ref{field-notation}) and the field-free 
one~(\ref{free-notation}) which is done for the states studied here in 
table \ref{table-correspondence}.

\begin{table*}[ht]
\begin{center}
\caption[Correspondence table between field and field-free notation]
{Correspondence table between the field~(\ref{field-notation}) and 
field-free (\ref{free-notation}) notation for the doubly excited 
helium atom for $B=0$. The field-free energy values are taken from the literature.
 \label{table-correspondence}}

%\scriptsize
%\footnotesize
\small
\begin{tabular}{*{2}{l} l|l *{2}{l}}
\multicolumn{6}{l}{  }\\
\hline
\multicolumn{2}{l}{$S=0:$\qquad \quad field-free}& \multicolumn{1}{c|}{field} &\multicolumn{2}{l}{$S=1:$\qquad \quad field-free} &\multicolumn{1}{c}{field}\\
\multicolumn{1}{c|}{energy}&\multicolumn{1}{l|}{\scriptsize{$L^\Pi\,_N(K,T)_n^A$}} &\multicolumn{1}{l|}{$_N\nu M^{\Pi_z}$} &\multicolumn{1}{c|}{energy}& \scriptsize{$L^\Pi\,_N(K,T)_n^A$}&\multicolumn{1}{|l}{$_N\nu M^{\Pi_z}$}\\
\hline 

0.77786764$^{\,a}$ &$^1\tn{S}^\tn{e}\, _2(1,0)_2^+$ &$_2 1^10^+$      &0.76049239$^{\,d}$ &$^3\tn{P}^\tn{o}\,_2(1,0)_2^+$ & $_2 1^30^-$\\
0.7019457$^{\,c}$  &$^1\tn{D}^\tn{e}\, _2(1,0)_2^+$ &$_2 2^10^+$      &                   &                               & $_2 1^3(\pm1)^+$\\
                   &                                &$_2 1^1(\pm1)^-$ &0.7105002$^{\,b}$  &$^3\tn{P}^\tn{e}\,_2(0,1)_2^+$ & $_2 1^30^+$\\
                   &                                &$_2 1^1(\pm2)^+$ &                   &                               & $_2 1^3(\pm1)^-$\\
0.69313495$^{\,d}$ &$^1\tn{P}^\tn{o}\, _2(0,1)_2^+$ &$_2 1^10^-$      &0.6025775$^{\,a}$  &$^3\tn{S}^\tn{e}\,_2(1,0)_3^-$ & $_2 2^30^+$\\
                   &                                &$_2 1^1(\pm1)^+$ &0.5846723$^{\,d}$  &$^3\tn{P}^\tn{o}\,_2(1,0)_3^+$ & $_2 2^30^-$\\
0.62192725$^{\,a}$ &$^1\tn{S}^\tn{e}\,_2(\tn{-1}.0)_2^+$&$_2 3^10^+$  &                   &                               & $_2 2^3(\pm1)^+$\\
0.5970738$^{\,d}$  &$^1\tn{P}^\tn{o}\,_2(1,0)_3^-$  &$_2 2^10^-$      &0.58378427$^{\,c}$ &$^3\tn{D}^\tn{e}\,_2(1,0)_3^-$ & $_2 3^30^+$\\
                   &                                &$_2 2^1(\pm1)^+$ &                   &                               & $_2 2^3(\pm1)^-$\\ 
0.58989468$^{\,a}$ &$^1\tn{S}^\tn{e}\, _2(1,0)_3^+$ &$_2 4^10^+$      &                   &                               & $_2 1^3(\pm2)^+$\\
0.58025$^{\,b}$    &$^1\tn{P}^\tn{e} \,_2(0,1)_3^-$ &$_2 5^10^+$      &0.57903099$^{\,d}$ &$^3\tn{P}^\tn{o}\,_2(0,1)_3^-$ & $_2 3^30^-$\\
                   &                                &$_2 1^1(\pm1)^-$ &                   &                               & $_2 3^3(\pm1)^+$\\
0.569221$^{\,c}$   &$^1\tn{D}^\tn{e}\, _2(1,0)_3^+$ &$_2 6^10^+$      &0.56781$^{\,b}$    &$^3\tn{P}^\tn{e}\,_2(0,1)_3^+$ & $_2 4^30^+$\\
                   &                                &$_2 3^1(\pm1)^-$ &                   &                               & $_2 3^3(\pm1)^-$\\ 
                   &                                &$_2 2^1(\pm2)^+$ &0.5662$^{\,b}$     &$^3\tn{F}^\tn{o}\,_2(1,0)_3^0$ & $_2 4^30^-$\\
0.564085$^{\,d}$   &$^1\tn{P}^\tn{o}\, _2(0,1)_3^+$ &$_2 3^10^-$      &                   &                               & $_2 4^3(\pm1)^+$\\
                   &                                &$_2 3^1(\pm1)^+$ &                   &                               & $_2 1^3(\pm2)^-$\\
0.5638$^{\,b}$     &$^1\tn{D}^\tn{o}\, _2(1,1)_2^+$ &$_2 4^10^-$      &                   &                               & $_2 1^3(\pm3)^+$\\
                   &                                &$_2 4^1(\pm1)^+$ &0.560687$^{\,c}$   &$^3\tn{D}^\tn{e}\,_2(0,1)_3^0$ & $_2 5^30^+$\\
                   &                                &$_2 1^1(\pm2)^-$ &                   &                               & $_2 4^3(\pm1)^-$\\
                   &                                &                 &                   &                               & $_2 2^3(\pm2)^+$\\

\hline
\multicolumn{6}{l}{$^a$ B\"urgers \cite{burgers},  $^b$ Lindroth \cite{lindroth},  $^c$ Ho  \cite{hoD}, $^d$ Ho \cite{hoP}}\\ 
\end{tabular}
\end{center}

\end{table*}

Table~\ref{Tabelle-0+B0N2-Genauigkeit} provides the relative accuracy of our 
calculated energies $E$ and decay widths $\Gamma$ for the energetically lowest 
resonances in the absence of the field. The comparison was performed with the
corresponding high precision results available in the 
literature~\cite{burgers,lindroth,hoD,hoP}. Generally it has to be remarked 
that for the states considered here, the absolute values of the energies 
are significantly larger than the values of the corresponding widths. 
The accuracy of our results is consequently much higher for the energies 
compared to the widths. Except for the states $3\, ^10^+$ and 
$4\, ^3(\pm1)^+$ all calculated energies possess a relative accuracy 
in the range of $7.5\cdot10^{-4}$ to $2\cdot10^{-5}$.

\begin{table}[ht]
\footnotesize

\begin{center}
\caption{The relative accuracy of the energetically lowest resonances of doubly excited helium for $B=0$ in the subspaces $\Pi_z=+1$ and $M=0,\pm1$ and $\pm2$ with $S=0,1$, $S_z=0$ and $N=2$ are presented. The accuracies are ordered by their degree of excitation $\nu$ within a given symmetry subspace (cf. table~\ref{table-correspondence}). % is such that they follow from top to bottom an increasing energy of the corresponding resonance with the given symmetry.
 \label{TableB0N2} \label{Tabelle-0+B0N2-Genauigkeit}}
\footnotesize

\begin{tabular}{*{2}{c}{r}}%*{2}{l}*{2}{c}}

%\hline 
$ \ \nu \ $&%$-E$&$\Gamma/2$&$-E_\tn{Lit}$&$\Gamma_\tn{Lit}/2$
$\Delta E$&\multicolumn{1}{c}{$\Delta \Gamma$}\\

\hline
\multicolumn{3}{c}{$M=0$  $S=0$}\\

1&%0.777813&0.002245930&0.77786764$^{\,a}$&0.002270653&
7.07E-05&1.09E-02$^{\,a}$\\
2&%0.701565&0.001121346&0.7019457$^{\,c}$&0.00118311&
5.41E-04&5.22E-02$^{\,b}$\\
3&%0.620618&0.000032051&0.62192725$^{\,a}$&0.000107818&
2.10E-03&7.03E-01$^{\,a}$\\
4&%0.589853&0.000643545&0.58989468$^{\,a}$&0.000681239&
7.12E-05&5.53E-02$^{\,a}$\\

\hline 
\multicolumn{3}{c}{$M=0$  $S=1$}\\

1&%0.710411&0.000000014&0.7105002$^{\,b}$&&
1.25E-04&$^{\,c}$\\
2&%0.602549&0.000000617&0.6025775$^{\,a}$&0.000003325&
4.81E-05&8.13E-01$^{\,a}$\\
3&%0.583770&&0.58378427$^{\,c}$&0.000000014&
2.39E-05&$^{\,b}$\\
\hline
\multicolumn{3}{c}{$M=\pm1$  $S=0$ }\\ 
 
1&%0.6927721&0.0007081&0.69313495$^{\,d}$&0.00068665&
5.24E-04&3.12E-02$^{\,d}$\\
2&%0.5970492&0.0000030&0.5970738$^{\,d}$&0.00000193&
4.19E-05&4.83E-01$^{\,d}$\\
3&%0.5639548&0.0001702&0.564085$^{\,d}$&0.00015053&
2.30E-04&1.30E-01$^{\,d}$\\
4&%0.5633774&0.0000117&0.5638$^{\,b}$&&
7.50E-04&$^{\,c}$\\
\hline
\multicolumn{3}{c}{$M=\pm1$  $S=1$ }\\

1&%0.76044&0.0001504&0.76049239$^{\,d}$&0.000149431&
6.84E-05&6.59E-03$^{\,d}$\\
2&%0.58465055&0.0000293&0.5846723$^{\,d}$&0.0000411&
3.59E-05&2.88E-01$^{\,d}$\\
3&%0.57899873&0.0000044&0.57903099$^{\,d}$&0.00000095&
5.53E-05&3.45E\.00$^{\,d}$\\
4&%0.56568&0.0000436&0.5662$^{\,b}$&0.000012&
9.18E-04&2.72E\.00$^{\,c}$\\

\hline 
\multicolumn{3}{c}{$M=\pm2$ $S=0$ }\\

1&%0.701566&0.00112135&0.7019457$^{\,c}$&0.00118311&
5.41E-04&5.22E-02$^{\,b}$\\
2&%0.569126&0.00026138&0.569221$^{\,c}$&0.0002775&
1.67E-04&5.81E-02$^{\,b}$\\

\hline 
%\multicolumn{3}{l}{  }\\
\multicolumn{3}{l}{\footnotesize $^a$ B\"urgers \cite{burgers},$^b$ Ho  \cite{hoD},}\\ 
\multicolumn{3}{l}{\footnotesize $^c$ Lindroth \cite{lindroth}, $^d$ Ho \cite{hoP}}\\ 
%\multicolumn{3}{l}{}\\ 
\end{tabular}
\end{center}

\end{table}

Obviously our approach is not best-suited to describe the resonances 
in the absence of the external field i.e. cannot compete with the best methods 
available in the literature based e.g. on a Hylleraas basis. 
However, our focus is to provide first results on resonances in strong 
magnetic fields and in this case our anisotropic Gaussian basis set is 
certainly superior to existing field-free methods that cannot efficiently 
describe the pronounced anisotropy present in a strong field. This reflects 
itself also in the different symmetries for $B=0$ and $B\neq0$.

\subsection{Resonances in a magnetic field}

\subsubsection{Dependence of the resonance energies on $M$ and $S_z$}

The Zeeman term of the Hamiltonian~(\ref{hamiltonian1}) depends linearly on $L_z$ 
and remains unchanged when the complex rotation is applied. 
Using eigenfunctions of $L_z$, as it is done here, one arrives at
\begin{eqnarray}
\langle i \vert T_\tn{Zeeman}(M)\vert j\rangle & = 
\langle i \vert\frac{1}{2}L_zB \vert j\rangle = & \frac{1}{2}MBS_{ij} , \nonumber  
\end{eqnarray}
with $S_{ij}$ being the overlap matrix of the two-particle states 
$\vert i\rangle$ and $\vert j\rangle$. The relation between two matrix 
elements associated with different signs of the magnetic quantum number 
$M$ reads as follows 
\begin{eqnarray}
\quad \langle i\vert T_\tn{Zeeman}(-M)\vert j\rangle & = &  
\langle i \vert T_\tn{Zeeman}(M)\vert j\rangle -MBS_{ij}. \nonumber
\end{eqnarray}
i.e. we obtain a shift. A similar argument holds for $S_z$,
\begin{eqnarray}
\langle i \vert T_\tn{Spin}(S_z)\vert j\rangle & = 
\langle i \vert S_zB \vert j\rangle = & S_zBS_{ij} . \nonumber  
\end{eqnarray}
In combination, the relation for the dependence of the complex energy 
$W$ on $M$ and $S_z$ can be put in the following way:
\begin{eqnarray}\label{MS-relation}
 W(M,S_z) & = &  W(-M,S_z=0) + MB + S_zB.
\end{eqnarray}
In what follows, we only present results for states belonging to 
subspaces with $M\leq0$ and $S_z=0$. All energies of subspaces with  
$M > 0$ and $S_z \neq 0$ are just shifted by a real energy term and 
can be calculated with the help of (\ref{MS-relation}). 
Note that the resonance widths do not depend on $S_z$ and the sign of $M$.

\subsubsection{Resonance energies}\label{subsubsec:real-energies}

The results of our calculations include the energies of the energetically 
lowest resonance states of the subspaces with $M=0,-1,-2$, $\Pi_z=+1$, $S_z=0$ 
and $S=0,1$. The values of the magnetic field strength are spread over a 
logarithmic scale covering the range $B=0 - 100$ a.u. ($B= 0 - 2.35\cdot 10^7$ T). 
The resulting data are given in the tables \ref{table-S0p1} to \ref{table-S2p1}.

The minimal energy of a single free electron ($m\leq0$) in a magnetic field 
is $E_{\tn{e}^-}(B)=B/2$ (cf. Landau energy~\cite{landau}). 
Therefore, the minimal energy of two non-interacting free electrons 
that possess no kinetic energy along the magnetic field is 
$E_{2\tn{e}^-}(B)=B$, which corresponds to the double ionization threshold 
of helium with a fixed nucleus in a magnetic field. In order to get the 
binding energies $E_\tn{bin}(B)$ of the helium states, this energy 
$E_{2\tn{e}^-}(B)$ has to be subtracted from the total energies: 
\begin{equation}\label{He-binding-energies}
E_\tn{bin}(B)=E(B)-E_{2\tn{e}^-}(B)=E(B)-B.
\end{equation}

\begin{figure}[tb]
\begin{center}
\includegraphics[width=0.49\textwidth%,height=0.35\textheight,angle=0
]{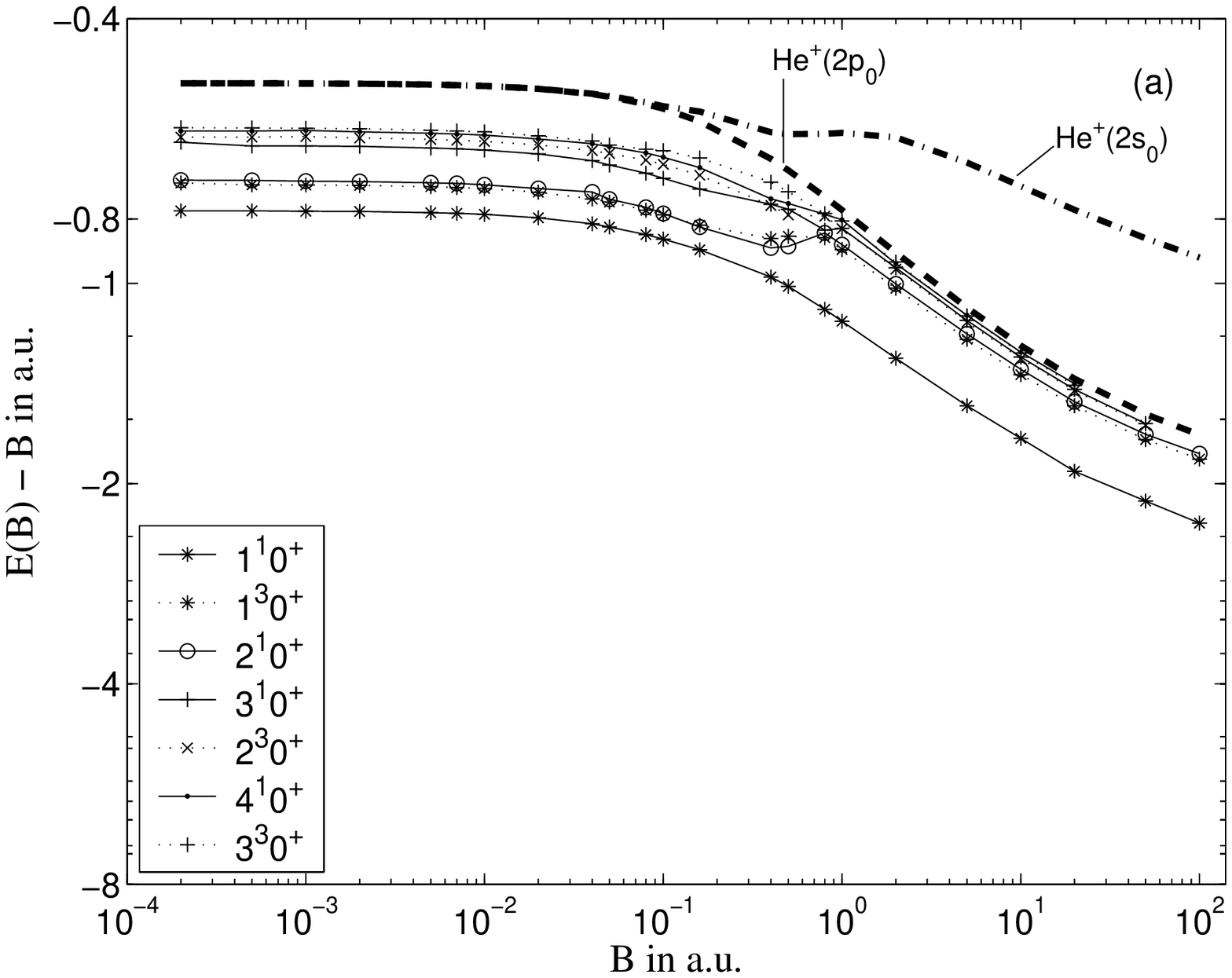}
\includegraphics[width=0.49\textwidth%,height=0.35\textheight,angle=0
]{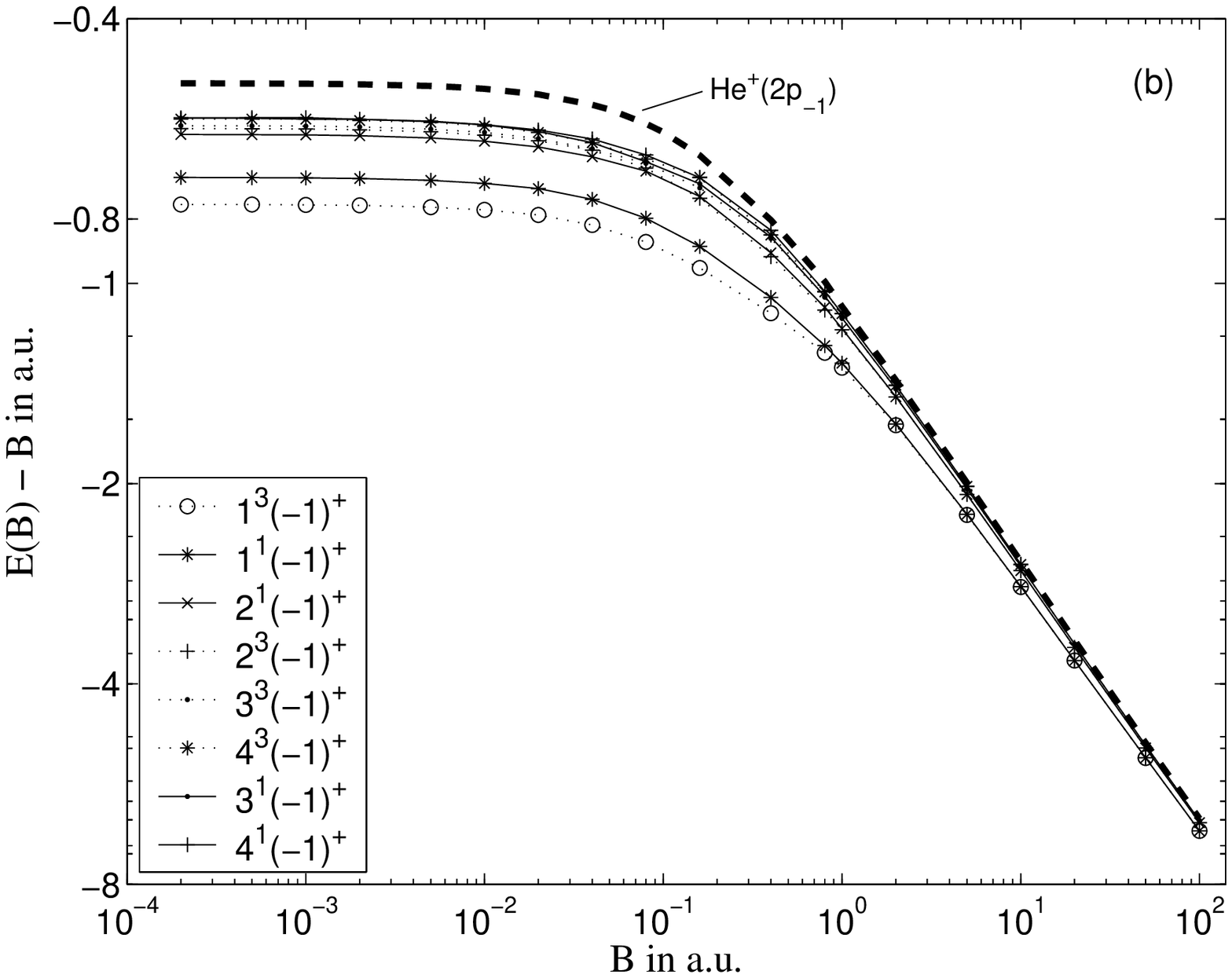}
\caption{Binding energies $E_\tn{bin}(B)=E(B)-B$ of doubly excited helium in a magnetic field range $0.0002\leq B\leq100$. The dependence on the magnetic field is shown for the energetically lowest resonances of the subspaces $M=0$ $\Pi_z=+1$ (a) and $M=-1$ $\Pi_z=+1$ (b).  Solid lines (---) stand for singlet  and dotted lines ($\cdots$) for triplet states. Additionally, appropriate binding energies of He$^+$ ions are given (\textbf{- -}) which correspond to upper thresholds.
\label{figure-binding}}
\end{center}
\end{figure}
Figure~\ref{figure-binding} shows the binding energies for the resonance 
states belonging to the subspaces $M=0$ (a) and $M=-1$ (b). 
The absolute values of the binding energies 
increase with increasing field strength for all resonant 
states considered here. The rate of increase, however, is larger for the 
resonances belonging to the symmetry subspace $M=-1$. Additionally, 
the binding energies of the He$^+$ ions $E_{\tn{He}^+}(B)$ emerging from 
the field-free quantum number $n=2$ and magnetic quantum numbers $m=0$ 
(a) and $m=-1$ (b) are shown. The total energies $E_\tn{He$^+$,tot}(B)$ 
of He$^+$ in a magnetic field can be obtained via a scaling relation~\cite{ruder} 
from the precisely known energies of hydrogen in a magnetic field~\cite{ruder,kravchenko}. 
In order to get the binding energies $E_{\tn{He}^+}(B)$ of the He$^+$ ion 
(similar as in (\ref{He-binding-energies})), $E_{\tn{e}^-}(B)$ has to be 
subtracted from the total energies $E_\tn{He$^+$,tot}(B)$:
\begin{equation}\label{He+-binding-energies}
E_{\tn{He}^+}(B)=E_\tn{He$^+$,tot}(B)-E_{\tn{e}^-}(B)=E_\tn{He$^+$,tot}(B)-B/2.\nonumber
\end{equation}
One observes that the binding energies of He$^+$(2p$_0$) and He$^+$(2p$_{-1}$) 
correspond to upper thresholds for $E_\tn{bin}(B)$ of the calculated states 
with $M=0$ (a) and $M=-1$ (b), respectively\;\footnote{The quantum number $n$ 
and the angular momentum $l$ are no good quantum numbers for $B>0$. 
However, they could be used for finite field strength as labels for 
states having these quantum numbers for $B=0$~\cite{ruder}.}, 
and they show the same qualitative behavior as the energies $E_\tn{bin}(B)$ 
of our resonances with increasing field strength. This can be interpreted in 
the following way: as a rough approximation one, of the electrons of the 
DE helium occupies the corresponding orbital of the He$^+$ ion mentioned above. 
In order to analyze the behavior of the second electron we 
subtract from the binding energies of the DE helium atoms the corresponding 
He$^+$ energy $E_{\tn{He}^+}(B)$:
\begin{equation}
E_2(M,B)=E_\tn{bin}(M,B) - E_{\tn{He}^+}(\tn{2p}_M,B).\nonumber
\end{equation}
\begin{figure}[tb]

\begin{center}
\includegraphics[width=0.49\textwidth%,height=0.35\textheight,angle=0
]{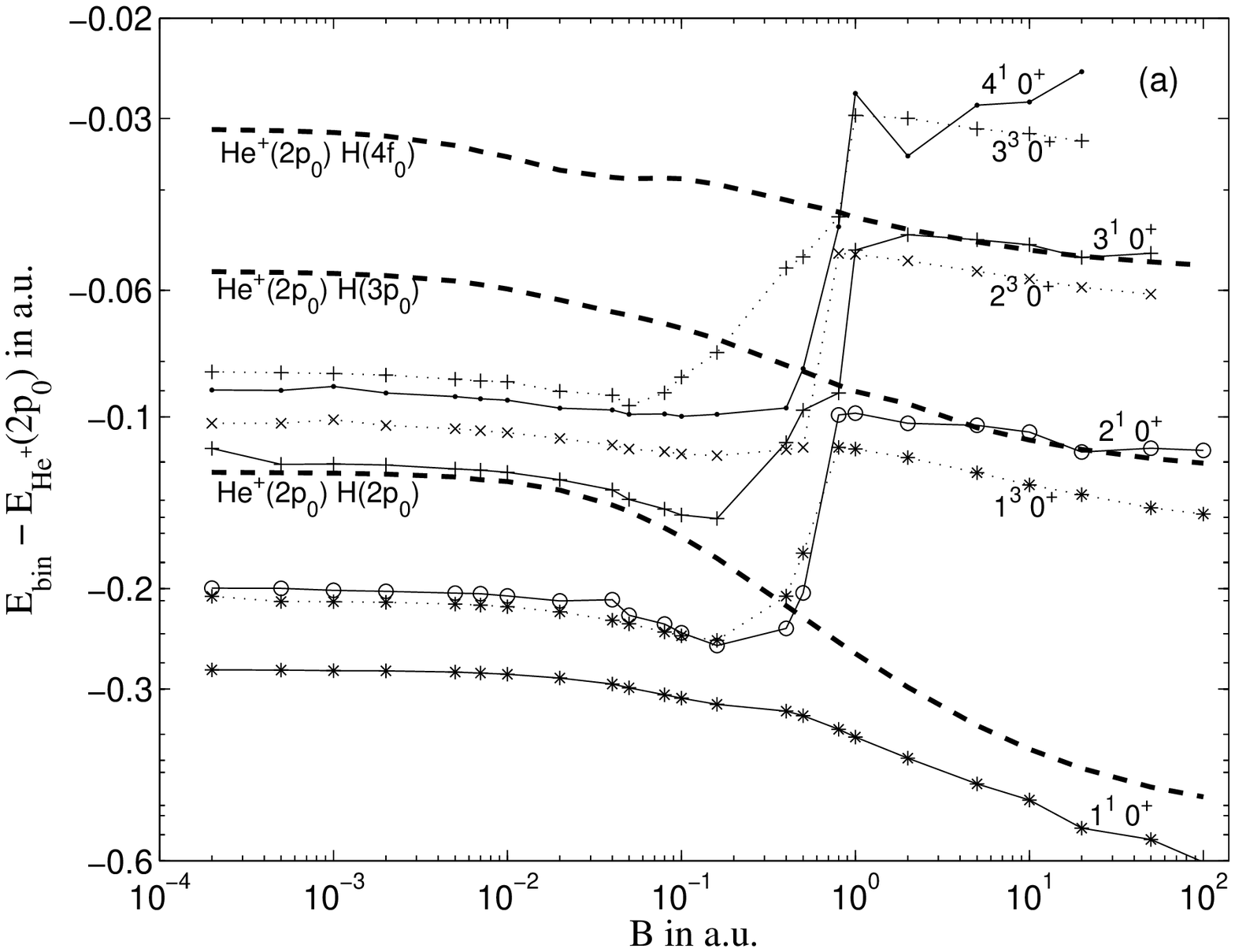}
\includegraphics[width=0.49\textwidth%,height=0.35\textheight,angle=0
]{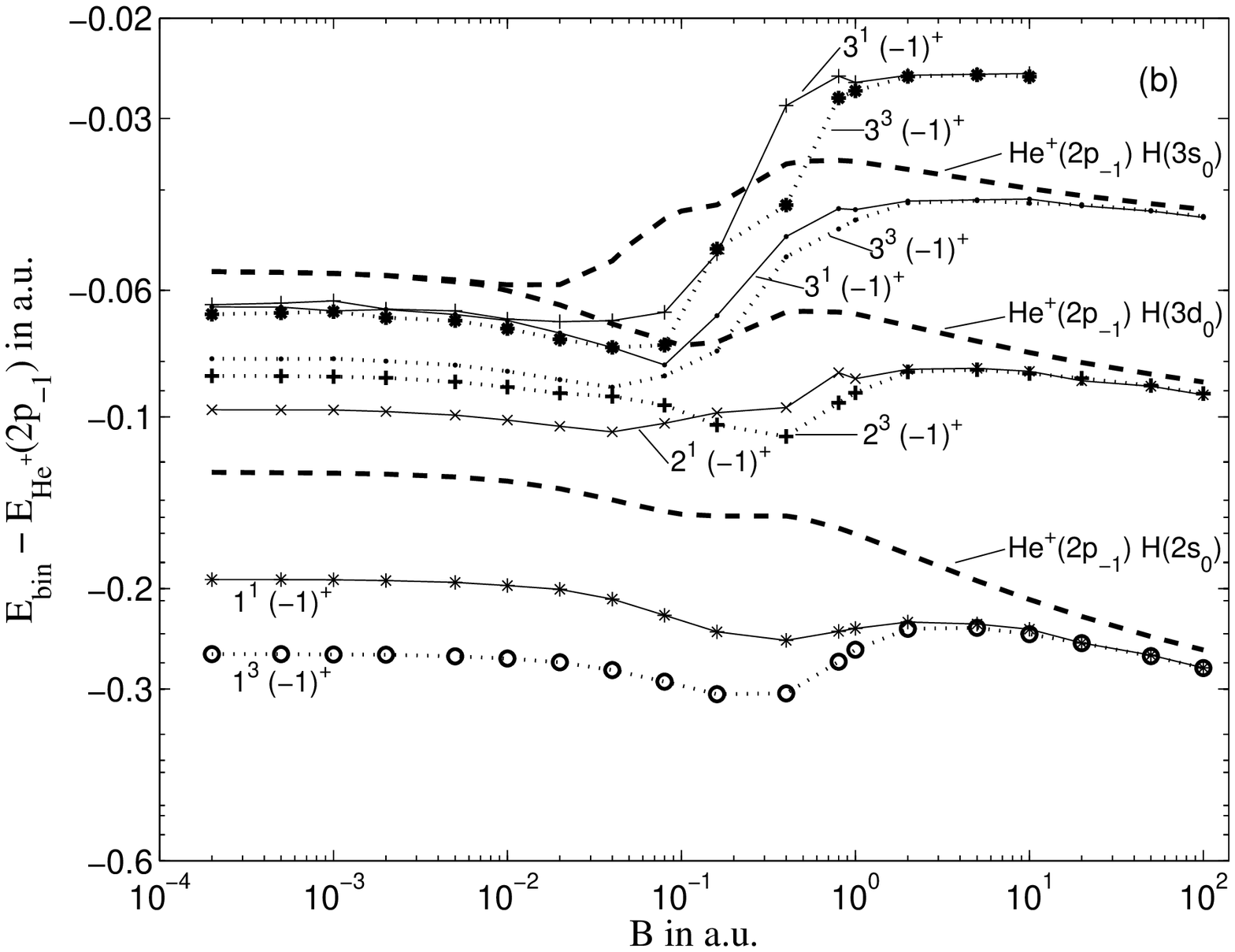}
\caption{Differences of the binding energies of the doubly excited helium atom and the He$^+$ ion $E_\tn{bin}(B)-E_{\tn{He}^+}(B)$ in a magnetic field range $0.0002\leq B\leq100$. The dependence on the magnetic field is shown for the energetically lowest states of the subspaces $M=0$ $\Pi_z=+1$ (a) and $M=-1$ $\Pi_z=+1$ (b). Solid lines (---) stand for singlet and dotted lines ($\cdots$) for triplet states. Additionally to $E_\tn{bin}(B)$, approximate energies are given (\textbf{- -}) which are a sum of the binding energies of a He$^+$ ion and a hydrogen atom (see text for details).
\label{figure-2.electron}}

\end{center}
\end{figure}
Corresponding results are shown in figure~\ref{figure-2.electron}. 
One observes a qualitatively different behavior of the energy levels 
according to three different regimes: I) low field regime $0\leq B\leq 10^{-2}$, 
II) intermediate field regime $10^{-2}<B\leq 1$ and III) high field regime 
$1<B\leq 10^{2}$. Especially in I) but also in III) there is a comparatively 
weak dependence on the magnetic field. On the other hand in II) a 
reorganization of the energy levels takes place being most pronounced in 
figure~\ref{figure-2.electron}a. This behavior is also known for hydrogen 
and singly excited helium in a magnetic field.
In both subfigures three additional energies are shown (\textbf{- -}). 
These approximate energies $E_\tn{app}(B)$ are the sum of the binding energy 
$E_{\tn{He}^+(\tn{2p}\,m_\tn{He$^+$})}(B)$ of the helium ion He$^+$(2p$_0$) (a), He$^+$(2p$_{-1}$) (b) 
and the energetically lowest binding energies  $E_{\tn{H}(nlm_\tn{H})}(B)$ 
of the hydrogen atom with $n\geq 2$: 
\begin{equation}
E_\tn{app}(B)=E_{\tn{He}^+(\tn{2p}m_{\tn{He}^+})}(B)+E_{\tn{H}(nlm_\tn{H})}(B).\nonumber
\end{equation}
The combination of the individual quantum numbers is in agreement with 
the total quantum numbers of the considered subspace: 
\begin{equation}
M=m_{\tn{He}^+}+m_\tn{H} \quad \tn{and} \quad 
\Pi_z=(\pi_{\tn{He}^+})_z\cdot(\pi_\tn{H})_z.\nonumber
\end{equation}
In these approximate energies, no interaction between the two electrons 
is taken into account, and a complete screening of one charge unit of the 
nucleus by the inner electron is assumed. 
Considering the subspace $M=-1$ in figure~\ref{figure-2.electron}b one 
can see that in the high field regime III) the energies $E_\tn{app}(B)$ 
come energetically close to the calculated resonance energies and show 
the same qualitative behavior. In this regime also a degeneracy of the singlet (---) 
and triplet ($\cdots$) states takes place (only $S_z=0$ is considered here). 
These features become more pronounced with increasing field strength and higher 
degree of excitation (logarithmic scale!). Let us concentrate as an example on the 
approximation $E_\tn{app}$(\small{He$^+$(2p$_{-1}$)\,H(2s$_0$)}) to the energies 
of the singlet and triplet ground states $1 ^1(-1)^+$ and $1 ^3(-1)^+$, respectively. 
It obviously significantly
underestimates the magnitude of their binding energies in the field regime 
I) and II) where both states are also energetically significantly separated. 
This is a sign for the strong overlap of the two electrons of the DE helium. 
This overlap decreases in the region III) and the electrons become spatially 
separated. Therefore, it is possible in III) to speak of an "inner" He$^+$-like 
electron, which is located close to the nucleus and screens one charge unit and 
of an "outer" H-like electron at a distance noticing only one effective positive charge. 

The above-observed behavior can be illustrated by using an independent-particle 
picture. All one-electron states except the \emph{tightly bound states} are elongated 
along the magnetic field axis and squeezed perpendicular to it with increasing 
magnetic field strength. Tightly bound states (1s$_{0}$, 2p$_{-1}$, 3d$_{-2}$, $\ldots$) 
are confined parallel to the magnetic field axis as well, and their binding energy rises 
strongly with increasing field strength~\cite{ruder}. In case that one of the electrons 
occupies a tightly bound orbital, the overlap of two electrons steadily decreases with 
increasing magnetic field strength. Thus the tightly bound electron is located close to 
the nucleus screening one charge, while the second electron stays mainly near the 
$z$-axis at a distance to the core and behaves therefore similar to an electron of a hydrogen atom. 
This scenario applies to the subspace $M=-1$ $\Pi_z=+1$ since 2p$_{-1}$ is a 
tightly bound orbital and consequently all other possible orbitals of the 
second electron are not tightly bound.

In the subspace $M=0$ in figure~\ref{figure-2.electron}a no degeneracy of the 
energies of the singlet and triplet states occurs in the presence of a strong field. 
However, the singlet states can be well approximated by the corresponding $E_\tn{app}(B)$. 
An exception to this rule is the ground state $1 ^1(0)^+$, which is approximated 
only qualitatively by the binding energies of He$^+$(2p$_0$) and H(2p$_0$), which 
corresponds to an intrashell state. In an intrashell state with both electrons 
approximately in the same orbital, it is expected that the assumption of 
weakly interacting electrons fails. Of course, there is no triplet partner for the 
ground state which is approximately described by the symmetric (2p$_0)^2$ 
configuration. In contrast to the subspace $M=-1$ the resonances of the subspace 
$M=0$ contain no (dominating) configurations that include a tightly bound orbital. 
As a consequence also the "inner" He$^+$-like electron is elongated along the 
$z$-axis and a noticeable overlap of the electronic orbitals remains 
even in the strong field regime III). 

The strong increase of some energies in the regime $0.1\leq B\leq 1$ 
(see figures~\ref{figure-2.electron}a and \ref{figure-2.electron}b) 
can be partly modeled with two-electron configurations that consist of  an 
electronic wave function of a helium ion He$^+$(2s$_0$) %(or even He$^+$(2p$_{+1}$)) 
and a hydrogen atom and are not shown in figure~\ref{figure-2.electron}. 
This effect results from the increasing energetic gap between the 
binding energy of He$^+$(2s$_0$) 
and He$^+$(2p$_0$) which can be seen in figure~\ref{figure-binding}a. 
Note that the approximations applied in the strong magnetic 
field regime III) do not hold in the regimes I) and II) and particularly not 
in the field-free situation~\cite{lin}.

\begin{figure}[tb]
\begin{center}
\includegraphics[width=0.49\textwidth%,height=0.35\textheight,angle=0
]{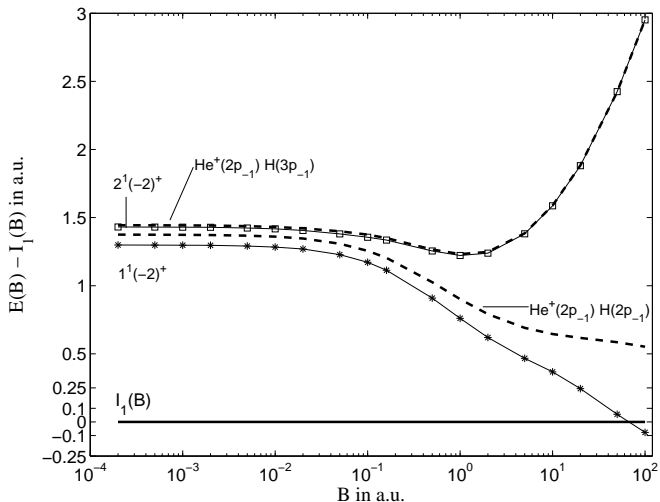}
\caption{Energetic difference  $E(B)-I_1(B)$ of DE helium energies to the first ionization threshold $I_1(B)$. The energies of the two energetically lowest singlet states of the subspace $M=-2$ $\Pi_z=+1$ in a magnetic field range $0.0002\leq B\leq100$ are presented. Additionally, approximate energies (\textbf{- -}) are given which are a sum of the binding energies of a He$^+$ ion and a hydrogen atom.
\label{figure-2+}}
\end{center}
\end{figure}

As a next step we aim at investigating the energetically lowest DE helium state 
in a strong magnetic field. Therefore, we searched for the DE helium state with 
the strongest binding energy in the field range III). The latter is dominated 
by the configuration He$^+$(2p$_{-1}$) H(2p$_{-1}$)~\cite{ruder}. Consequently 
we calculated the energetically lowest singlet states of the subspace $M=-2$ and $\Pi_z=+1$.

In figure~\ref{figure-2+} the energetic distances of the two energetically lowest 
singlet states to the first ionization threshold $I_1(B)=E_{\tn{He}^+(1\tn{s}_0)}(B)+B$ 
are shown. The energies of both states show a weak dependence on the field 
strength for $B<0.01$. For $B>0.01$ the energetic difference of the 
energetically lowest resonance state $1^1(-2)^+$ to $I_1(B)$ decreases with 
increasing field strength. A crossover of the first ionization threshold seems to 
take place for $B\approx65$ a.u. This would correspond to a stabilization of the 
resonance $1^1(-2)^+$ and a conversion to a bound state for a magnetic field 
$B>65$ since the autoionization process is not possible anymore. The energetically 
lowest resonance is approximated only qualitatively by  the configuration 
involving the non-interacting orbitals He$^+$(2p$_{-1}$) and H(2p$_{-1}$), 
for it is an intrashell state which consists of two tightly bound orbitals.
We note that the above considerations assume a spin singlet for both
the resonance and the threshold state, which is indeed not the energetically
lowest state in the strong field regime.

The second singlet state $2^1(-2)^+$ is approximated well by the configuration 
involving the orbitals He$^+$(2p$_{-1}$), H(3p$_{-1}$) and for $B>1$ there is 
hardly a difference visible on the scale of figure~\ref{figure-2+}. This can be 
explained by applying the same argument as for the subspace $M=-1$ where all 
calculated states also consist approximately of a tightly bound and a non-tightly 
bound orbital. However, the energetic difference to $I_1(B)$ increases strongly 
in the high field regime and no crossing of $I_1(B)$ is possible. 

Using the results which have been obtained so far one can relate the 
doubly excited helium states in a magnetic field denoted by the 
spectroscopic field notation~(\ref{field-notation}) in the limit of a high 
field strength to the corresponding dominant independent particle configurations 
of a helium ion and hydrogen electron. This has been done in table~\ref{table_approx}. 
In addition, the corresponding correlation quantum numbers~(\ref{free-notation}) 
for a vanishing magnetic field $B=0$ are given.

\begin{table}[tb]
\begin{center}
\caption[]{Correspondence between the spectroscopic field and field-free notation for vanishing magnetic field. Additionally, for a strong magnetic field the corresponding approximate configurations of a He$^+$ and a H electron are presented. Configuration marked by $\dagger$ are intrashell states which quantitatively do not approximate the energies well.
\label{table_approx}}
\small
\begin{tabular}{ccc}
\hline
\footnotesize
\footnotesize field not.&\footnotesize field-free not.&\footnotesize app. config.\\
\footnotesize $B\geq0$&\footnotesize $B=0$&$B\rightarrow \infty$\\

\hline
$1^10^+$&$^1\tn{S}^\tn{g}\, _2(1,0)_2^+$&2p$_0$2p$_0\,\dagger$ \\
$2^10^+$&$^1\tn{D}^\tn{g}\, _2(1,0)_2^+$&2p$_0$3p$_0$\\
$3^10^+$&$^1\tn{S}^\tn{g}\, _2(-1,0)_2^+$&2p$_0$4f$_0$\\
$4^10^+$&$^1\tn{S}^\tn{g}\, _2(1,0)_3^+$&2p$_0$4p$_0$\\
\hline
$1^30^+$&$^3\tn{P}^\tn{g}\, _2(0,1)_2^+$&2p$_0$3p$_0$\\
$2^30^+$&$^3\tn{S}^\tn{g}\, _2(1,0)_3^-$&2p$_0$4f$_0$\\
$3^30^+$&$^3\tn{D}^\tn{g}\, _2(1,0)_3^-$&2p$_0$4p$_0$\\
\hline
$1^1(-1)^+$&$^1\tn{P}^\tn{u} \, _2(0,1)_2^+$&2p$_{-1}$2s$_0$\\
$2^1(-1)^+$&$^1\tn{P}^\tn{u} \, _2(1,0)_3^-$&2p$_{-1}$3d$_0$\\
$3^1(-1)^+$&$^1\tn{P}^\tn{u} \, _2(0,1)_3^+$&2p$_{-1}$3s$_0$\\
$4^1(-1)^+$&$^1\tn{D}^\tn{u} \, _2(1,1)_2^+$&2p$_{-1}$4d$_0$\\
\hline
$1^3(-1)^+$&$^3\tn{P}^\tn{u} \, _2(1,0)_2^+$&2p$_{-1}$2s$_0$\\
$2^3(-1)^+$&$^3\tn{P}^\tn{u} \, _2(1,0)_3^+$&2p$_{-1}$3d$_0$\\
$3^3(-1)^+$&$^3\tn{P}^\tn{u} \, _2(0,1)_3^-$&2p$_{-1}$3s$_0$\\
$4^3(-1)^+$&$^3\tn{F}^\tn{u} \, _2(1,0)_3^0$&2p$_{-1}$4d$_0$\\
\hline
$1^1(-2)^+$&$^1\tn{D}^\tn{g}\, _2(1,0)_2^+$&2p$_{-1}$2p$_{-1}\,\dagger$\\
$2^1(-2)^+$&$^1\tn{D}^\tn{g}\, _2(1,0)_3^+$&2p$_{-1}$3p$_{-1}$
\end{tabular}
\end{center}
\end{table}

\subsubsection{Resonance widths}

Parallel to the real energies dealt with in subsection \ref{subsubsec:real-energies}, 
the corresponding imaginary parts $\Gamma$ have been calculated, too. 
$\Gamma$ is interpreted as the width of the resonance and therefore as the 
inverse lifetime. Typically, the absolute values of the widths are by orders of 
magnitude smaller than the real energies. Our method allows for the same absolute 
accuracy for both the real and the imaginary part. This leads to a significantly 
lower relative accuracy of the resonance widths compared to the real energies. The widths 
themselves vary by orders of magnitudes. Therefore, it is most natural to inspect 
the behavior of those widths more closely which possess a large absolute value 
($\Gamma/2 \geq 0.0001$).
\begin{figure}[tb]

\begin{center}
\includegraphics[width=0.5\textwidth%,height=0.35\textheight,angle=0
]{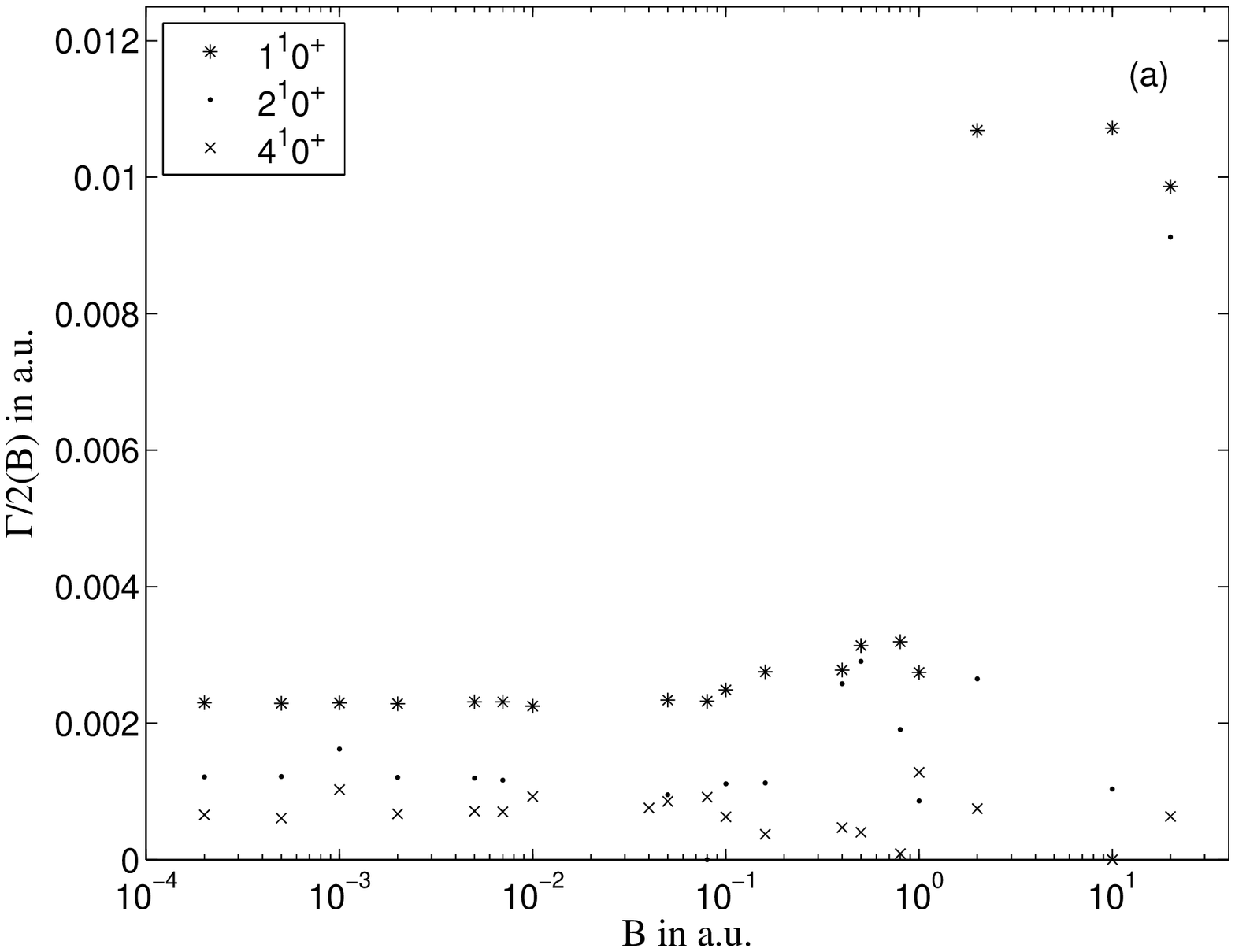}
\includegraphics[width=0.485\textwidth%,height=0.35\textheight,angle=0
]{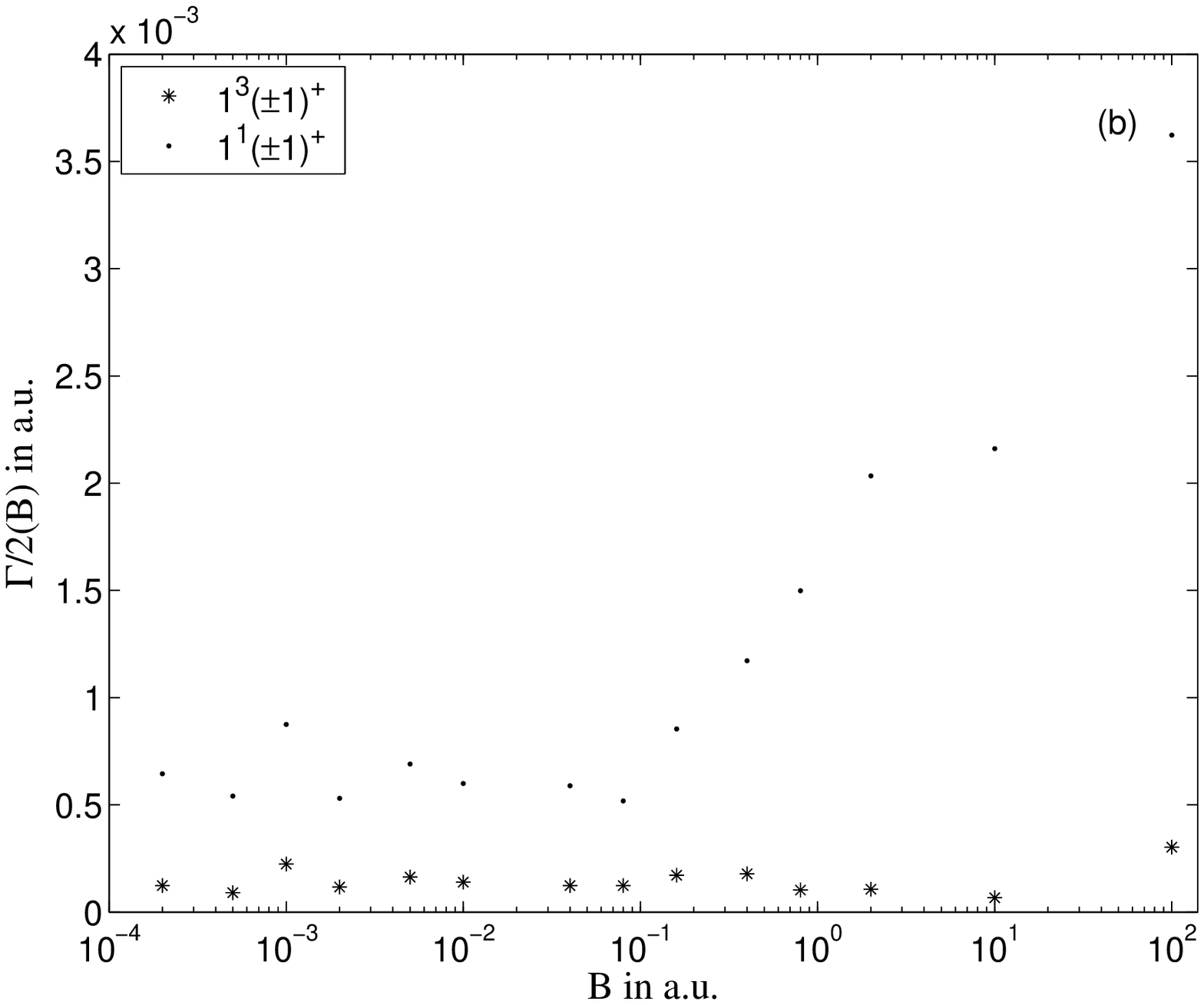}
\caption{Resonance widths $\Gamma$ of the DE helium atom in a magnetic field range $0.0002\leq B\leq 100$. The widths of three singlet states with $M=0$ $\Pi_z=+1$ (a) and of the two ground states with $M=\pm1$ $\Pi_z=+1$ (b) are plotted. 
\label{figure-widths}}

\end{center}
\end{figure}
These include three singlet states of the $M=0$ $\Pi_z=+1$ subspace 
$1^10^+$, $2^10^+$ and $4^10^+$, which are shown in figure~\ref{figure-widths}a. 
Additionally, the widths of the two energetically lowest resonance states of the $M=-1$ $\Pi_z=+1$ 
subspace, $1^1(-1)^+$ and $1^3(-1)^+$, are presented in figure~\ref{figure-widths}b. 
The corresponding data can be found in the tables \ref{table-S0p1}, \ref{table-S0p2}, 
\ref{table-S1p1} and \ref{table-T1p1}.

For $B<0.05$ just a weak dependence on the magnetic field can be observed for 
the three singlet states in figure~\ref{figure-widths}a. With further increasing 
field strength, the widths start to fluctuate, although  this fluctuation could 
possibly be a consequence of the finite numerical accuracy. The width of the 
state  $1^10^+$ increases in the regime $1<B<2$ approximately by a factor of $4$.

In figure~\ref{figure-widths}b we compare a singlet and a triplet state 
otherwise possessing the same quantum numbers. In figure~\ref{figure-2.electron}b 
we have seen that the energies of these states become degenerate for $B>2$. 
For $B<0.1$ their widths fluctuate around a mean value of 
$7 \cdot 10^{-4}$ ($1^1(-1)^+$) and $1.5 \cdot 10^{-4}$ ($1^3(-1)^+$), respectively. 
For $B>0.1$ the width of the triplet state keeps fluctuating, but the width of the 
singlet states increases with the magnetic field strength $B$. 
This means that both states become energetically degenerate for strong magnetic 
fields, but the behavior of the corresponding resonance widths is 
different in this field range.

%%%%%%%%%%%%%%%%%%%%%%%%%%%%%%%%%%%%%%
%
\section{Summary and Outlook}\label{sec:summary}
%
%%%%%%%%%%%%%%%%%%%%%%%%%%%%%%%%%%%%%%

We have investigated the energies and lifetimes of the resonance states of a 
doubly excited helium atom exposed to a magnetic field. The nonrelativistic, 
infinite-mass Hamiltonian is characterized by the symmetries associated 
with the four quantum numbers: $M$, $\Pi_z$, $\bm S^2$ and $S_z$. 
Our computational approach consists of the complex scaling method applied 
to a full configuration-interaction scheme based on nonlinearly optimized 
anisotropic Gaussian orbitals.

In order to show the behavior of the DE helium states denoted by the 
spectroscopic field notation~(\ref{field-notation}) in the limits of weak
and strong magnetic field strength, we have presented for vanishing magnetic 
field their corresponding correlation quantum numbers~(\ref{free-notation}) 
and for a strong magnetic field their corresponding independent-particle 
configurations of a helium ion and hydrogen electron.

In total the energies of 17 DE states have been calculated. Their behavior 
can be subdivided into three magnetic field ranges: the weak ($0\leq B\leq10^{-2}$), 
the intermediate ($10^{-2}<B\leq 1$) and the strong ($1<B\leq 10^2$) field regime. 
With increasing field strength we observe the transition from a weak to a strong 
dependence on the magnetic field which is accompanied by a reorganization of the energies of the 
resonances especially in the intermediate field regime.

In the strong field regime, a clustering of the calculated energies into different 
branches according to their magnetic quantum numbers takes place. Furthermore, 
the electron-electron interaction becomes less important with increasing field 
strength, being more pronounced for the states of the subspace $M=-1$ than for 
those of $M=0$. As a result, a crude description of the resonance energies by the 
sum of the energies of a He$^+$ ion and a hydrogen atom is possible. 
An exception to this rule are the intrashell states.
The DE helium state with the lowest energy ($S_z=0$) is the state $1^1(-2)^+$ 
of the subspace $M=-2$. This resonance state can probably be stabilized by a 
magnetic field, for its calculated energy becomes smaller than the one-electron 
ionization threshold for a field strength of $B>65$. 

The behavior of the widths of five calculated resonances with great 
absolute value ($\Gamma\gtrsim10^{-4}$) has been inspected. The dependence of the 
widths on the magnetic field for low field strengths $B<0.1$ is weak. The 
values of the widths are fluctuating around their mean values, which could 
be due to the finite numerical accuracy of our method. For field strengths 
$B>1$ the widths of two states, namely $1^10^+$ and $1^1(-1)^+$, significantly 
increase while the widths of the other investigated states still fluctuate 
with an increasing amplitude.

Having presented the first data on the energies of the resonances of DE helium 
in a magnetic field, several issues still have to be addressed 
in future works. In order to get synthetic spectra of astronomical objects 
like MWD, the number of calculated states should be increased including 
also states with $\Pi_z=-1$. Additionally, magnetic field dependent oscillator 
strengths and electromagnetic-transition probabilities have to be studied. 
On the other hand, a closer look at the resonance widths would be desirable in 
order to gain a deeper understanding of the role of electron-electron 
interactions and the autoionization process under the influence of a magnetic field. 
An increased accuracy of our data would be a prerequisite for this.

\section{Acknowledgments}\label{sec:acknowledgments}
We are grateful to H.-D. Meyer for enlightening discussions.
%\vspace{12cm}

%\input{include_N20p-Sing1}

\begin{table*}[p]
%\begin{center}
\caption[]{Calculated energies $E$ and resonance widths $\Gamma$ of resonances of the subspace $^1(0)^+$ and $N=2$ for different magnetic field strengths $B$:  1$^1(0)^+$ and 2$^1(0)^+$ (field-free $^1\tn{S}^e\,_2(1,0)_2^+$ and $^1\tn{D}^e\,_2(1,0)_2^+$).
\label{table-S0p1}}

\small
%\footnotesize
\begin{tabular}{*{3}{|c}|*{2}{|c}|}
\hline

&\multicolumn{2}{c| |}{ 1$^1(0)^+$} & \multicolumn{2}{c|}{ 2$^1(0)^+$}\\ 

%\hline
$B$&$-E$&$\Gamma/2$&$-E$&$\Gamma/2$\\
\hline\hline
0&0.77786$\,^a$&0.002271&0.70195$\,^b$&0.001181\\
\hline
0&0.77781&0.002246&0.70157&0.001121\\
0.0002&0.77762&0.002299&0.69948&0.001211\\
0.0005&0.77763&0.002291&0.6995&0.001217\\
0.001&0.77802&0.002296&0.70093&0.001619\\
0.002&0.77781&0.002283&0.70116&0.001206\\
0.005&0.77776&0.002312&0.70112&0.001193\\
0.007&0.77772&0.002311&0.70064&0.001163\\
0.01&0.77758&0.002249&0.70098&0.000523\\
0.02&0.77691&0.002683&0.70006&0.00225\\
0.04&0.77348&0.001094&0.68867&0.000155\\
0.05&0.7726&0.002337&0.69688&0.000952\\
0.08&0.76462&0.002319&0.68842&\\
0.1&0.75772&0.002487&0.68534&0.001109\\
0.16&0.72982&0.002751&0.66255&0.001122\\
0.4&0.57765&0.002779&0.48461&0.002578\\
0.5&0.51019&0.003137&0.37951&0.002904\\
0.8&0.29357&0.003192&0.03998&0.001906\\
1&0.13985&0.002744&-0.12582&0.00086\\
2&-0.70432&0.01068&-0.99833&0.002647\\
5&-3.47217&0.005766&-3.80865&0.000267\\
10&-8.29029&0.010718&-8.65359&0.001035\\
20&-18.0834&0.009865&-18.4943&0.00912\\
50&-47.8771&0.007483&-48.3142&0.002379\\
100&-97.7081&&-98.1969&\\

\hline
\multicolumn{5}{l}{$^a$ B\"urgers \cite{burgers}, $^b$ Ho \cite{hoD}} 
\end{tabular}

%\end{center}
\end{table*}
\begin{table*}[p]
%\begin{center}

\caption[]{Calculated energies $E$ and resonance widths $\Gamma$ of resonances of the subspace $^1(0)^+$ and $N=2$ for different magnetic field strengths $B$:  3$^1(0)^+$ and 4$^1(0)^+$ (field-free $^1\tn{S}^e\,_2(-1,0)_2^+$ and $^1\tn{S}^e\,_2(1,0)_3^+$).
\label{table-S0p2}}

\small
%\footnotesize
\begin{tabular}{*{1}{|c}*{1}{|c}|*{2}{|c}|}
\hline

&\multicolumn{1}{c| |}{3$^1(0)^+$}& \multicolumn{2}{c|}{4$^1(0)^+$}\\ 

%\hline
$B$&$-E$&$-E$&$\Gamma/2$\\

\hline\hline
0&0.62193$\,^a$&0.58989$\,^a$&0.000681\\
\hline
0&0.62062&0.58985&0.000644\\
0.0002&0.61347&0.58965&0.000655\\
0.0005&0.62085&0.58960&0.000609\\
0.001&0.62044&0.58794&0.001025\\
0.002&0.62044&0.58984&0.000670\\
0.005&0.62085&0.58960&0.000710\\
0.007&0.62048&0.58941&0.000699\\
0.01&0.62009&0.58849&0.000924\\
0.02&0.61873&0.58637&0.000377\\
0.04&0.61373&0.57662&0.000755\\
0.05&0.61368&0.57308&0.000852\\
0.08&0.60268&0.55647&0.000914\\
0.1&0.59495&0.54606&0.000625\\
0.16&0.56150&0.50973&0.000372\\
0.4&0.36064&0.34611&0.000468\\
0.5&0.27335&0.25830&0.000402\\
0.8&0.03153&-0.01283&0.000086\\
1&-0.17336&-0.19726&0.001278\\
2&-1.05301&-1.06606&0.000748\\
5&-3.86318&-3.88367&\\
10&-8.71004&-8.73188&\\
20&-18.55700&-18.58471&0.000631\\
50&-48.37604&-48.41477&0.000049\\
100&-98.19686&-98.29527&\\
\hline
\multicolumn{4}{l}{$^a$ B\"urgers \cite{burgers}} 
\end{tabular}

%\end{center}
\end{table*}
\begin{table*}[p]
%\begin{center}

\caption[]{Calculated energies $E$ of resonances of the subspace $^3(0)^+$ and $N=2$ for different magnetic field strengths $B$:  1$^3(0)^+$, 2$^3(0)^+$ and 3$^3(0)^+$ (field-free $^3\tn{P}^e\,_2(0,1)_2^+$, $^3\tn{S}^e\,_2(-1,0)_3^-$ and $^3\tn{D}^e\,_2(1,0)_3^-$).
\label{table-T0p1}}

\small
%\footnotesize
\begin{tabular}{*{2}{|c}|*{1}{|c}|*{1}{|c}|}
\hline

&\multicolumn{1}{c| |}{1$^3(0)^+$}&\multicolumn{1}{c| |}{2$^3(0)^+$}& \multicolumn{1}{c|}{3$^3(0)^+$}\\ 

%\hline
$B$&$-E$&$-E$&$-E$\\

\hline\hline
0&0.71050$\,^a$&0.60258$\,^b$&0.58378$\,^c$ \\
\hline
0&0.71041&0.60255&0.58359 \\
0.0002&0.70621&0.60250&0.58331 \\
0.0005&0.71039&0.60237&0.58344 \\
0.001&0.71045&0.60064&0.58344 \\
0.002&0.71036&0.60254&0.58345 \\
0.005&0.71039&0.60237&0.58344 \\
0.007&0.71033&0.60222&0.58298 \\
0.01&0.71014&0.60157&0.58176 \\
0.02&0.70952&0.59899&0.58001 \\
0.04&0.70672&0.59146&0.57100 \\
0.05&0.70465&0.58797&0.56960  \\
0.08&0.69598&0.57264&0.54839 \\
0.1&0.68825&0.56251&0.53147\\
0.16&0.65724&0.52765&0.48792 \\
0.4&0.45585&0.36377&0.30446 \\
0.5&0.34928&0.28915&0.22842 \\
0.8&0.05382&-0.00755&-0.01468 \\
1&-0.11049&-0.8276&-0.19474 \\
2&-0.98307&-1.04769&-1.07099 \\
5&-3.78678&-3.85651&-3.88082 \\
10&-8.62833&-8.70260&-8.72806 \\
20&-18.47258&-18.55026&-18.57670 \\
50&-48.28332&-48.36675&-48.40350 \\
100&-98.16338&-98.25919&-98.28200 \\

\hline
\multicolumn{4}{l}{$^a$ Lindroth \cite{lindroth}, $^b$ B\"urgers \cite{burgers}, $^c$ Ho \cite{hoD}} 
\end{tabular}

%\end{center}
\end{table*}
\begin{table*}[p]
%\begin{center}
\caption[]{Calculated energies $E$ and resonance widths $\Gamma$ of resonances of the subspace $^1(-1)^+$ and $N=2$ for different magnetic field strengths $B$:  1$^1(-1)^+$ 2$^1(-1)^+$, 3$^1(-1)^+$ and 4$^1(-1)^+$ (field-free $^1\tn{P}^o\,_2(0,1)_2^+$, $^1\tn{P}^o\,_2(1,0)_3^-$,  $^1\tn{P}^o\,_2(0,1)_3^+$ and $^1\tn{D}^o\,_2(1,1)_2^+$).
\label{table-S1p1}}

\small
%\footnotesize
\begin{tabular}{*{3}{|c}|*{1}{|c}|*{1}{|c}|*{1}{|c}|}
\hline

&\multicolumn{2}{c| |}{ 1$^1(-1)^+$}&\multicolumn{1}{c| |}{2$^1(-1)^+$} &\multicolumn{1}{c| |}{3$^1(-1)^+$}&\multicolumn{1}{c|}{4$^1(-1)^+$}\\ 

%\hline
$B$&$-E$&$\Gamma/2$&$-E$&$-E$&$-E$\\
\hline\hline

0&0.69313$\,^a$&0.000687&0.59707$\,^a$&0.56408$\,^a$&0.56338$\,^b$ \\
\hline
0&0.69277&0.000708&0.59705&0.56395&0.56338 \\
0.0002&0.69293&0.000645&0.59716&0.56413&0.56356 \\
0.0005&0.69286&0.000541&0.59724&0.56420&0.56317 \\
0.001&0.69306&0.000875&0.59723&0.56518&0.56260 \\
0.002&0.69359&0.000530&0.59783&0.56482&0.56470 \\
0.005&0.69499&0.000691&0.59929&0.56611&0.56517 \\
0.01&0.69746&0.000599&0.60123&0.56766&0.56732 \\
0.02&0.70032&0.001664&0.60359&0.57098&0.56781 \\
0.04&0.70755&0.000589&0.60505&0.57442&0.56664 \\
0.08&0.71809&0.000518&0.59790&0.57637&0.56085 \\
0.1&0.71239&0.000749&0.59139&0.56711&0.54302 \\
0.16&0.71991&0.000854&0.58014&0.54826&0.53355 \\
0.4&0.64988&0.001143&0.49967&0.45167&0.43185 \\
0.8&0.43997&0.001498&0.28585&0.24529&0.22741 \\
1&0.31808&0.000575&0.16893&0.12654&0.10915 \\
2&-0.37324&0.002034&-0.51960&-0.56024&-0.57691 \\
5&-2.77600&0.000682&-2.92460&-2.96510&-2.98170 \\
10&-7.14520&0.002161&-7.29770&-7.33940&-7.35590 \\
20&-16.31200&0.001250&-16.47400&-16.51800&-16.58900 \\
50&-44.83400&0.001380&-45.00700&-45.05200&\\
100&-93.35100&0.003623&-93.53500&-93.58200& \\
\hline
\multicolumn{6}{l}{$^a$ Ho \cite{hoP}, $^b$ Lindroth \cite{lindroth}} 
\end{tabular}

%\end{center}
\end{table*}
\begin{table*}[p]
%\begin{center}

\caption[]{Calculated energies $E$ and resonance widths $\Gamma$ of resonances of the subspace $^3(-1)^+$ and $N=2$ for different magnetic field strengths $B$:  1$^3(-1)^+$ 2$^3(-1)^+$, 3$^3(-1)^+$ and 4$^3(-1)^+$ (field-free $^3\tn{P}^o\,_2(1,0)_2^+$, $^3\tn{P}^o\,_2(1,0)_3^+$, $^3\tn{P}^o\,_2(0,1)_3^-$ and $^3\tn{F}^o\,_2(1,0)_3^0$).
\label{table-T1p1}}

\small
%\footnotesize

\begin{tabular}{*{3}{|c}|*{1}{|c}|*{1}{|c}|*{1}{|c}|}
\hline

&\multicolumn{2}{c| |}{ 1$^3(-1)^+$}&\multicolumn{1}{c| |}{2$^3(-1)^+$} &\multicolumn{1}{c| |}{3$^3(-1)^+$}&\multicolumn{1}{c|}{4$^3(-1)^+$}\\ 

%\hline
$B$&$-E$&$\Gamma/2$&$-E$&$-E$&$-E$\\
\hline\hline

0&0.76049$\,^a$&0.000149&0.58465$\,^a$&0.57900$\,^a$&0.56568$\,^b$\\
\hline
0&0.76044&0.000150&0.58465&0.57900&0.56568 \\
0.0002&0.76055&0.000123&0.58475&0.57908&0.56608 \\
0.0005&0.76059&0.000090&0.58484&0.57913&0.56571 \\
0.001&0.76092&0.000224&0.58498&0.57905&0.56541 \\
0.002&0.76121&0.000117&0.58542&0.57985&0.56701 \\
0.005&0.76284&0.000163&0.58682&0.58115&0.56777 \\
0.01&0.76516&0.000140&0.58861&0.58313&0.56994 \\
0.02&0.76911&0.000611&0.59056&0.58572&0.57277 \\
0.04&0.77664&0.000123&0.59090&0.58756&0.57433 \\
0.08&0.78635&0.000123&0.59067&0.58004&0.57017 \\
0.1&0.78735&0.000161&0.58934&0.56246&0.54601 \\
0.16&0.78813&0.000172&0.58511&0.55843&0.53253 \\
0.4&0.70867&0.000172&0.51165&0.45580&0.44592 \\
0.8&0.47094&0.000102&0.29665&0.24898&0.22976 \\
1&0.33918& &0.17395&0.12839&0.11003 \\
2&-0.36680&0.000106&-0.51867&-0.55994&-0.57678 \\
5&-2.77240&&-2.92410&-2.96500&-2.98160 \\
10&-7.14050&0.000067&-7.29700&-7.33870&-7.35560 \\
20&-16.3110& &-16.47500&-16.51800&-16.59300 \\
50&-44.8330& &-45.0080&-45.0520&-45.070000 \\
100&-93.3510&0.000302&-93.5350&-93.58200& \\
\hline
\multicolumn{4}{l}{$^a$ Ho \cite{hoP}, $^b$ Lindroth \cite{lindroth}} 
\end{tabular}
%\end{center}
\end{table*}
\begin{table*}[p]
%\begin{center}
\caption[]{Calculated energies $E$ and resonance widths $\Gamma$ of resonances of the subspace $^1(-2)^+$ and $N=2$ for different magnetic field strengths $B$:  1$^1(-2)^+$ and 2$^1(-2)^+$ (field-free $^1\tn{D}^e\,_2(1,0)_2^+$ and $^1\tn{D}^e\,_2(1,0)_3^+$).
\label{table-S2p1}}
\small
%\footnotesize
\begin{tabular}{*{3}{|c}|*{2}{|c}|}

\hline

&\multicolumn{2}{c| |}{ 1$^1(-2)^+$} & \multicolumn{2}{c|}{ 2$^1(-2)^+$}\\

$B$&$-E_\tn{res}$&$\Gamma/2$&$E_\tn{res}-B$&$E_\tn{Aug}$\\
\hline\hline

0&0.70195$\,^a$&0.001183&0.56922$\,^a$&0.000278\\
\hline
0&0.70157&0.001121&0.56913&0.000261\\
0.0005&0.70203&0.001122&0.56960&0.000261\\
0.001&0.70249&0.001125&0.57009&0.000264\\
0.002&0.70249&0.001125&0.57009&0.000264\\
0.005&0.70643&0.001052&0.57379&0.000249\\
0.01&0.71116&0.001105&0.57755&0.000270\\
0.02&0.72055&0.001152&0.58413&0.000283\\
0.05&0.74541&0.001196&0.59369&0.000308\\
0.1&0.77727&0.001020&0.59346&0.000433\\
0.16&0.80518&0.001557&0.58362&0.000615\\
0.5&0.82534&0.004077&0.47937&0.000573\\
1&0.68014&0.006971&0.21906&0.000840\\
2&0.16923&0.005071&-0.44999&0.000117\\
5&-1.92293&0.007648&-2.83753&0.001424\\
10&-5.97779&0.002951&-7.19591&0.000081\\
20&-14.78130& &-16.35960&0.000128\\
50&-42.51036& &-44.87839&\\
100&-90.36262& &-93.39070&0.000145\\

%-0,0768

\hline
\multicolumn{5}{l}{$^a$ Ho \cite{hoD}} 

\end{tabular}

%\end{center}
\end{table*}

\end{document}